\documentstyle[eqsecnum,preprint,aps]{revtex}

\def\footnoterule{\kern-3pt \hrule width \hsize \kern6.2pt}

\def\pmb#1{\setbox0=\hbox{$#1$}
\kern-.025em\copy0\kern-\wd0
\kern.05em\copy0\kern-\wd0
\kern-.025em\raise.0433em\box0 }
\tighten

\begin{document}

\title{Gauge Invariant Variables for\\
Spontaneously Broken ${\bf SU(2)}$ Gauge Theory\\
in the Spherical Ansatz}

\author{Edward Farhi\footnote[1]{farhi@mitlns.mit.edu.~~
This work is supported in part by funds provided by the
U.S.~Department of Energy (D.O.E.)
under cooperative agreement \#DF-FC02-94ER40818.}}
\smallskip
\address{Center for Theoretical Physics\\
Laboratory for Nuclear Science and\\
Department of Physics\\
Massachusetts Institute of Technology\\
Cambridge, MA\ \ 02139}

\author{Krishna Rajagopal\footnote[2]{rajagopal@huhepl.harvard.edu.~~
Junior Fellow, Harvard Society of Fellows.~~Research supported in part
by the Milton Fund of Harvard University and by
the National Science Foundation under grant PHY-92-18167.}}
\smallskip
\address{Lyman Laboratory of Physics\\
Harvard University\\
Cambridge, MA\ \ 02138}

\author{Robert Singleton, Jr.\footnote[3]{bobs@cthulu.bu.edu.~~Research
supported in part by the D.O.E. under contract \#DE-FG02-91ER40676 and
by an NSF Fellowship under grant number ASC-940031. }}
\smallskip
\address{Department of Physics\\
Boston University\\
Boston, MA\ \ 02215}

\maketitle

\setcounter{page}{0}
\thispagestyle{empty}

\vfill

\noindent CTP\#2415

\noindent HUTP-95-A007

\noindent BUHEP-95-4 \hfill Submitted to {\it Physical Review} {\bf D}

\noindent hep-ph/9503268 \hfill Typeset in REV\TeX
\eject

\vfill

\begin{abstract}

We describe classical solutions to the Minkowski space
equations of motion of $SU(2)$ gauge theory coupled
to a Higgs field in the spatial spherical ansatz. We show
how to reduce the equations to four equations for four gauge
invariant degrees of freedom which correspond to the massive
gauge bosons and the Higgs particle. The solutions typically
dissipate at very early and late times. To describe the solutions
at early and late times, we linearize and decouple the equations
of motion, all the while working only with gauge invariant variables.
We express the change in Higgs winding of a solution in terms
of gauge invariant variables.

\end{abstract}

\vfill

\eject

\baselineskip 24pt plus 2pt minus 2pt

\section{Introduction}
\label{sec1:level1}

In this paper, we develop techniques for solving
the Minkowski space classical equations of
motion for
$SU(2)$ gauge theory with spontaneous symmetry
breaking introduced via the Higgs mechanism.  (This
model is the standard electroweak theory without the
$U(1)$ gauge field and without the fermions.)
We work in the spatial spherical ansatz,
the equations for
which were first written down by Ratra and Yaffe \cite{W}.
Typical solutions to these equations of motion
represent inward and outward going spherical
shells.  The Ratra Yaffe equations are six equations
for six functions of $r$ and $t$.
Using the residual $U(1)$ gauge invariance which is present
in the spherical ansatz, we reduce this to
four equations for four gauge invariant functions
of $r$ and $t$  in a manner which guarantees
that Gauss's law is automatically satisfied at all time.
We linearize these equations and then
find new variables in which the linear equations decouple.
Typical solutions to the full equations are well
approximated by solutions to the linear equations
at early and late times. These solutions can be characterized
by their change in Higgs winding, which we write in terms
of gauge invariant variables.
We hope that the gauge invariant variables we introduce
will be useful to others studying classical solutions
in the spherical ansatz\cite{bobclaudio}.

In previous work done with V. V. Khoze \cite{FKRS}, we
considered $SU(2)$ gauge theory with no Higgs field.
There, we found that in the spherical ansatz
solutions have the property that in the far
past and far future they can be described
(after multiplying by the appropriate power of $r$)
as spherical
shells which propagate without distortion.
Here, we include the
Higgs field.  The classical equations of motion
are no longer scale invariant, and
have the dispersion characteristic of
wave equations for massive fields.  Thus, the
solutions we consider here are qualitatively
different than those considered in Ref. \cite{FKRS}.

\vskip .3in
\section{The Spherical Ansatz}
\label{sec2}

In this section,
we review the spherical ansatz \cite{W} for $SU(2)$
gauge theory with a Higgs field.
The action for this theory is
  \begin{equation}
  S=\int d^4x\,  \left\{ -\frac{1}{2} {\rm Tr}F^{\mu\nu}F_{\mu\nu}  -
  \frac{1}{2}{\rm Tr}\,\left(D^\mu\Phi\right)^\dagger  D_\mu\Phi
  - \frac{\lambda}{4}
  \left({\rm Tr}\Phi^\dagger \Phi - v^2\right)^2
  \right\} \ ,
  \label{Action}
  \end{equation}
where
  \begin{eqnarray}
  F_{\mu\nu} &=& \partial_\mu A_\nu - \partial_\nu A_\mu - i g [A_\mu,A_\nu]
  \nonumber\\
  D_\mu \Phi &=& (\partial_\mu - i g A_\mu)\Phi
  \label{Fmunu}
  \end{eqnarray}
with $A_\mu = A_\mu^a \sigma^a/2$
and where the $2\times 2$ matrix $\Phi$ is related to the Higgs
doublet $\varphi$ by
\begin{equation}
\Phi({\bf x},t) = \left( \matrix{ \varphi_2^* & \varphi_1 \cr
-\varphi_1^* & \varphi_2 } \right) \ .
\label{higgsmatrix}
\end{equation}
Following Ref. \cite{W} we use the
metric
$ds^2 = -dt^2 + d{\bf x}^2$. Our definition of $v$ follows standard
conventions, and is $\sqrt{2}$ times the $v$ of Ref. \cite{W}.

The spherical ansatz is given by expressing the gauge field $A_\mu$ and
the Higgs field $\Phi$ in terms of six
real functions  $a_0\, ,\,a_1\, , \, \alpha\, , \, \gamma\, , \,
\mu\ {\rm and}\ \nu\ {\rm of}\ r\
{\rm and}\ t$:
\begin{eqnarray}
  A_0({\bf x},t) &=& \frac{1}{2 g} \, a_0(r,t){\pmb\sigma}\cdot{\bf \hat x}
  \nonumber\\
  A_i({\bf x},t) &=& \frac{1}{2 g} \, \big[a_1(r,t){\pmb\sigma}\cdot{\bf\hat
x}\hat
  x_i+\frac{\alpha(r,t)}{r}(\sigma_i-{\pmb\sigma}\cdot{\bf\hat x}\hat x_i)
  +\frac{\gamma(r,t)}{r}\epsilon_{ijk}\hat x_j\sigma_k\big]
  \nonumber\\
  \Phi({\bf x},t) &=& \frac{1}{g}  \,  [\mu(r,t) + i \nu(r,t)
  {\pmb\sigma}\cdot{\bf \hat x}] \ ,
  \label{SphAn}
  \end{eqnarray}
where ${\bf \hat x}$ is the unit three-vector in the radial direction
and ${\pmb\sigma}$ are the matrices.  For the four dimensional
fields to be regular at the origin, we require that $a_0$, $\alpha$,
$a_1-\alpha/r$, $\gamma/r$ and $\nu$ vanish as $r \to 0$. Under
a gauge transformation of the form
$\exp [ i \Omega(r,t) {\pmb\sigma}\cdot{\bf \hat x}/2 ]$ with
$\Omega(0,t)=0$, configurations in the
spherical ansatz remain in the spherical ansatz and continue to satisfy
the appropriate boundary conditions at the origin.  Thus, the $SU(2)$
gauge theory reduced to the spherical ansatz has a residual $U(1)$
gauge invariance.

In the spherical ansatz the action (\ref{Action})
takes the form  \cite{W}
  \begin{eqnarray}
  S = \frac{4\pi}{g^2}  \int dt\int^\infty_0dr \, && \bigg[-\frac{1}{4}
  r^2f^{\mu\nu}f_{\mu\nu}-(D^\mu\chi)^*D_\mu \chi
  - r^2 (D^\mu\phi)^*D_\mu\phi
  \nonumber\\
  && -\frac{1}{2 r^2}\left( ~ |\chi |^2-1\right)^2
  -\frac{1}{2}(|\chi|^2+1)|\phi|^2 -  {\rm Re}(i \chi^* \phi^2)
  \nonumber \\
  && -\frac{\lambda}{g^2}  \, r^2 \, \left(|\phi|^2 -
  \frac{g^2 v^2}{2}\right)^2 ~
  \bigg] \ ,
  \label{SphAction}
  \end{eqnarray}
where the indices now run over $0$ and $1$ and
\begin{mathletters}
\begin{eqnarray}
f_{\mu\nu}&=&\partial_\mu a_\nu - \partial_\nu a_\mu
\label{deffmunu}  \\
\chi &=&\alpha+i\Big(\gamma-1\Big)
\label{defchi}  \\
\phi &=& \mu+i \nu\
\label{defphi}  \\
D_\mu\chi &=& (\partial_\mu-i  \, a_\mu)\chi
\label{defDchi}  \\
D_\mu \phi&=& (\partial_\mu- \frac{i}{2}  \, a_\mu)\phi\ .
\label{defDphi}
\end{eqnarray}
\end{mathletters}
The notation is chosen to manifest the
$U(1)$ gauge invariance present in
the action (\ref{SphAction}).
The complex scalar fields $\chi$ and $\phi$ have $U(1)$  charges
of $1$ and $1/2$ respectively,
$a_\mu$ is the $U(1)$ gauge field,
$f_{\mu\nu}$ is the field strength,
and $D_\mu$ is the covariant derivative.
The indices are raised and lowered with the
$1+1$ dimensional metric $ds^2 = -dt^2+dr^2$.

The equations of motion for the reduced theory are
  \begin{mathletters}
  \begin{equation}
  -\partial^\mu(r^2f_{\mu\nu})=i\left[D_\nu \chi^*\chi-\chi^*D_\nu\chi \right]
+
  \frac{i}{2}\,   \, r^2 \left[D_\nu \phi^*\phi-\phi^*D_\nu\phi\right]
  \label{fEq}
  \end{equation}
  \begin{equation}
  \left[-D^2+\frac{1}{r^2}(|\chi|^2-1) + \frac{1}{2}\,  |\phi |^2 ~
\right]\chi=
  -\frac{i}{2}\,  \, \phi^2
  \label{chiEq}
  \end{equation}
  \begin{equation}
  \left[-D^\mu r^2 D_\mu+\frac{1}{2}(|\chi|^2+1) +
  \frac{2\lambda}{g^2}\,r^2 \left(|\phi|^2-\frac{g^2 v^2}{2}\right)
  \right] \phi= i \, \chi \phi^* \ .
  \label{phiEq}
  \end{equation}
  \label{gvarEqs}
  \end{mathletters}
The same equations are obtained either by varying the action (\ref{Action})
and then imposing the spherical ansatz or by varying the
action (\ref{SphAction}).

\vskip .3in

\section{Gauge Invariant Variables}
\label{sec3}

Equations (\ref{gvarEqs}) are six real equations
for the six degrees of freedom in $\chi$, $\phi$, and
$a_\mu$. However they can be reduced to four
equations for four gauge invariant variables as we now
show.
First, we write the complex fields $\chi$ and $\phi$
in polar form\footnote[1]{In Ref. \cite{FKRS} the phase of $\chi$ was called
$\varphi$.
In this paper we call it $\theta$.
}
  \begin{mathletters}
  \begin{equation}
  \chi(r,t)=-i\rho(r,t)\exp[i \theta(r,t)]
  \label{rhoDef}
  \end{equation}
  \begin{equation}
  \phi(r,t)=\sigma(r,t)\exp[i \eta(r,t)]\ .
  \label{sigmaDef}
\end{equation}
\label{polarDef}
\end{mathletters}
The variables $\rho$ and $\sigma$ are gauge invariant, and
the $r=0$ boundary conditions on $\alpha$, $\gamma$ and
$\nu$ imply
\begin{mathletters}
\label{onebc}
\begin{eqnarray}
\rho(0,t) &=& 1
\label{rhobc}\\
\theta(0,t) &=& 0
\label{firstthetabc}\\
\eta(0,t) &=& n(t) \pi {\rm ~when~}
\sigma(0,t)\neq 0 \ ,
\label{firstetabc}
\end{eqnarray}
\end{mathletters}
where $n(t)$ is an integer.
The boundary condition (\ref{firstthetabc}) should strictly be that
$\theta(0,t)$ is an integer multiple of $ 2\pi$.  However, since $\rho$
never vanishes at the origin, $\theta$ is constant in time there and
we have taken it to vanish.  On the other hand, $\eta$ can change
discontinuously at the origin by an integer multiple of $\pi$ at  times
when $\sigma(0,t)$ vanishes.

Since we are in $1 + 1$ dimensions the gauge invariant field
strength $f_{\mu\nu}$ must be proportional to $\epsilon_{\mu\nu}$
so we can define the gauge invariant variable $\psi$ via
  \begin{equation}
  r^2 f_{\mu\nu} = - 2 \epsilon_{\mu\nu}\psi~.
  \label{psiDef}
  \end{equation}
(Here $\epsilon_{01}=+1).$
Since the fields $\chi$ and $\phi$ have $U(1)$ charges
of  $1$  and $1/2$ respectively, the phase variable
  \begin{equation}
  \xi = \theta - 2 \eta
  \label{xiDef}
  \end{equation}
is also gauge invariant.
Note that $\xi$ is periodic with period $2\pi$.
Also note that
$\xi$ is not defined when either  $\rho$ or $\sigma$ vanishes.
We now write the equations of motion in terms
of the four gauge invariant variables $\rho$, $\sigma$, $\psi$ and $\xi$.

Using the definitions (\ref{polarDef}),
equation (\ref{fEq})
becomes
  \begin{mathletters}
  \begin{equation}
 \partial^\mu(r^2f_{\mu\nu})+2\rho^2(\partial_\nu \theta-a_\nu) +
   r^2 \sigma^2  (\partial_\nu\eta- \frac{1}{2}a_\nu) = 0 \ ,
  \label{Four}
  \end{equation}
while (\ref{chiEq}) becomes the two real equations
  \begin{equation}
  \partial_\mu \partial^\mu \rho- \rho(\partial^\mu \theta-a^\mu)
  (\partial_\mu \theta-a_\mu)  - \frac{1}{r^2}( \rho^2 -1)\rho
  - \frac{1}{2}\, \sigma^2 \rho=-\frac{1}{2}\, \sigma^2 \cos  \xi
  \label{Five}
  \end{equation}
and
  \begin{equation}
  \partial^\mu \left[\rho^2 (\partial_\mu \theta-a_\mu)\right]=
  \frac{1}{2} \, \rho \, \sigma^2 \, \sin \xi \ ,
  \label{Six}
  \end{equation}
where we have used the definition (\ref{xiDef}) of $\xi$.
Similarly,
(\ref{phiEq}) becomes the two real equations
  \begin{equation}
  \partial_\mu r^2 \partial^\mu \sigma -  r^2 \sigma \, (\partial^\mu \eta
  -\frac{1}{2}a^\mu)(\partial_\mu \eta-\frac{1}{2}a_\mu)
  - \frac{1}{2}(  \rho^2 +1)\sigma - \frac{2 \lambda}{g^2} \, r^2
  \left(\sigma^2- \frac{g^2v^2}{2}\right)\sigma
  =- \rho \sigma \cos \xi
  \label{Seven}
  \end{equation}
and
  \begin{equation}
  \partial^\mu \left[r^2 \sigma^2 (\partial_\mu \eta-\frac{1}{2}a_\mu)\right]=
  - \rho \sigma^2 \sin \xi \ .
  \label{Eight}
  \end{equation}
  \label{giOne}
  \end{mathletters}
Note that applying $\partial^\nu$ to (\ref{Four})
yields twice (\ref{Six}) plus
(\ref{Eight}).  We can thus view  (\ref{Eight}) as redundant
and drop it.

Now, using (\ref{psiDef}) in (\ref{Four}) we obtain
  \begin{equation}
  2\epsilon_{\mu\nu} \partial^\mu \psi = 2\rho^2\partial_\nu
  \theta + r^2 \sigma^2 \partial_\nu\eta - 2  a_\nu \, (\rho^2 +
  \frac{1}{4} r^2 \sigma^2) \ .
  \label{temp}
  \end{equation}
Solving (\ref{temp}) for $a_\nu$ and using $\xi=\theta - 2\eta$ we
find
  \begin{equation}
  a_\nu = 2 \partial_\nu\eta + \frac{  \rho^2\partial_\nu\xi -
  \epsilon_{\mu\nu}\partial^\mu\psi}{\rho^2 + \frac{1}{4}r^2\sigma^2} \ ,
  \label{anuSolve}
  \end{equation}
from which we obtain the gauge invariant combinations
  \begin{mathletters}
  \begin{equation}
  \partial_\nu \theta - a_\nu = \frac{\epsilon_{\mu\nu}\partial^\mu\psi
  + \frac{1}{4}  r^2 \sigma^2 \partial_\nu \xi}{\rho^2 +
\frac{1}{4}r^2\sigma^2}
  \label{AAA}
  \end{equation}
  \begin{equation}
  \partial_\nu \eta - \frac{1}{2}a_\nu = \frac{1}{2} \,
\frac{\epsilon_{\mu\nu}\partial^\mu\psi
  - \rho^2 \partial_\nu \xi}{\rho^2 + \frac{1}{4}r^2\sigma^2} \ .
  \label{BBB}
  \end{equation}
  \label{dma}
  \end{mathletters}
{}From (\ref{psiDef}) we have that
$\epsilon^{\mu\nu}\partial_\mu a_\nu = 2 \psi/r^2$,
which using (\ref{anuSolve}) gives
  \begin{mathletters}
  \begin{equation}
  \partial^\mu\left\{ \frac{\partial_\mu \psi +   \rho^2 \epsilon_{\mu\nu}
  \partial^\nu\xi}{\rho^2 + \frac{1}{4}r^2\sigma^2}\right\} - \frac{2 }{r^2} \,
\psi = 0  \ .
  \label{giPsi}
  \end{equation}
Substituting (\ref{dma}) into (\ref{Five}), (\ref{Six}),
and (\ref{Seven}), we get
  \begin{equation}
  \partial_\mu\partial^\mu\rho
  - \frac{\rho \, (\frac{1}{4}  r^2\sigma^2\partial_\mu\xi
  -\epsilon_{\mu\nu}\partial^\nu\psi )^2}{(\rho^2 + \frac{1}{4}r^2\sigma^2)^2}
  - \frac{1}{r^2}( \rho^2-1)\rho
  -\frac{1}{2} \rho \, \sigma^2 + \frac{1}{2}  \, \sigma^2\cos  \xi = 0
  \label{giRho}
  \end{equation}
  \begin{equation}
  \partial^\mu\left\{ \frac{\rho^2(\frac{1}{4}  r^2\sigma^2\partial_\mu\xi -
  \epsilon_{\mu\nu}\partial^\nu\psi)}{\rho^2 + \frac{1}{4}r^2\sigma^2}\right\}
  - \frac{1}{2}\rho\sigma^2\sin  \xi= 0
  \label{giPsiXiOne}
  \end{equation}
  \begin{equation}
  \partial_\mu r^2\partial^\mu\sigma - \frac{\frac{1}{4}r^2\sigma \, (
\rho^2\partial_\mu\xi
  +\epsilon_{\mu\nu}\partial^\nu\psi )^2}{(\rho^2 + \frac{1}{4}r^2\sigma^2)^2}
  - \frac{1}{2}( \rho^2+1)\sigma
  -\frac{2\lambda}{g^2}\, r^2 \left(\sigma^2-\frac{g^2v^2}{2}\right)\sigma
  +  \rho \sigma\cos \xi = 0 \ .
  \label{giSigma}
  \end{equation}
  \label{giEqs}
  \end{mathletters}

We have succeeded in casting the equations
of motion as four equations (\ref{giEqs}) for four gauge
invariant variables. To solve these equations we need
the boundary conditions at the origin.  The condition
$\rho(0,t)=1$ was already given in (\ref{rhobc}),
and  from (\ref{psiDef}) we have $\psi(0,t)=0$. From
(\ref{firstthetabc}),  (\ref{firstetabc}) and (\ref{xiDef})  we
find that when $\sigma$ does not vanish at the origin,
$\xi(0,t) = 0 \,{\rm mod}\, 2\pi $. When $\sigma$ vanishes
at the origin, $\xi$ is not defined there. The boundary
condition on $\sigma$ is determined by examining the
small-$r$ behavior of (\ref{giEqs}) and demanding that
solutions be regular at the origin. When  $\sigma$ is
non-zero at the origin, the condition $\partial_r \sigma(0,t) =0$
must be imposed, and when $\sigma$ vanishes at the
origin the constraint on $\partial_r \sigma(0,t)$ is
that it be non-zero and finite.
Equations (\ref{giEqs}) can now be
solved after specifying initial value data,
that is the values of $\rho\, , \, \dot\rho\, , \,
\sigma \, , \, \dot\sigma\, , \,
\psi \, , \, \dot\psi \, , \, \xi$ and $\dot\xi$ at some initial
time. With a solution in hand, if a gauge is chosen
then the gauge variant variables $\chi$, $\phi$
and $a_\mu$ can be determined using (\ref{dma}).
(For example, consider $a_0=0$ gauge and make
a time-independent gauge transformation such that
$\eta=0$ and therefore $\theta=\xi$ at time $t=0$.
Then, with the gauge invariant variables $\rho$, $\sigma$,
$\psi$, and $\xi$ known, the
$\nu=0$ components of equations (\ref{dma})
allow one to obtain $\theta$ and $\eta$ at all
times.  The $\nu=1$ component of either of equations
(\ref{dma}) can then be solved for $a_1$.)
Any initial value data
expressed in terms of gauge invariant variables
yields, upon choosing a gauge, initial value
data in terms of gauge variant variables which is
consistent with the Gauss's
law constraint.
This is because Gauss's law is the
$\nu=0$ component
of (\ref{temp}).

It is useful to consider the energy functional obtained from
the action (\ref{SphAction}).  Using Gauss's law, the energy
can be written
  \begin{eqnarray}
  \label{Energy}
  E = \frac{8\pi}{g^2} && \int_0^\infty dr \bigg[ ~
\frac{1}{2}(\partial_t\rho)^2 +
  \frac{1}{2}(\partial_r\rho)^2 + r^2\left( \frac{1}{2} \, (\partial_t\sigma)^2
+
  \frac{1}{2} \, (\partial_r\sigma)^2 \right) + \\
  \nonumber &&
  \frac{\frac{1}{4}r^2\sigma^2\rho^2}
  {\rho^2 + \frac{1}{4}r^2\sigma^2 }\left( \frac{1}{2} (\partial_t\xi)^2 +
  \frac{1}{2}(\partial_r\xi)^2\right) + \frac{1} {\rho^2 +
\frac{1}{4}r^2\sigma^2 }
  \left( \frac{1}{2} (\partial_t\psi)^2 +  \frac{1}{2}(\partial_r\psi)^2\right)
+ \\
  \nonumber &&
  \frac{\psi^2}{r^2} + \frac{1}{4 r^2}(\rho^2-1)^2 + \frac{1}{4}(\rho^2+1)
  \sigma^2 - \frac{1}{2} \, \rho\sigma^2\cos \xi + \frac{1}{2}\frac{\lambda}
  {g^2} \, r^2  \left(\sigma^2- \frac{g^2v^2}{2}\right)^2 ~ \bigg]  \ .
  \end{eqnarray}
{}From (\ref{Energy}) we see that in vacuum,  $\rho_{{\rm vac}}=1$,
$\sigma_{{\rm vac}}= g v/\sqrt{2}$,  $\psi_{{\rm vac}}=0$ and
$\xi_{{\rm vac}}=0 \,{\rm mod}\, 2\pi $.
Note that $\xi$ is periodic with period $2\pi$, and
there is in fact only one vacuum
configuration.
The familiar winding
number associated with vacua is not seen
in gauge invariant variables, because
$SU(2)$ vacuum configurations with different
winding numbers are gauge transforms of one
another, and correspond to a single
point in the gauge invariant configuration
space described by $\rho$, $\sigma$, $\psi$, and $\xi$.

Since $\xi$ is periodic, the reader may be curious whether
configurations in which $\xi$ winds by $2\pi n$ are topological
solitons.
Such finite energy configurations
can be constructed,
but there is no topological obstruction
to their unwinding.  These configurations
have been studied in some detail by Turok and
Zadrozny \cite{Turok}.
Consider a finite energy configuration
with $\psi=0$, $\rho=1$ and $\sigma=gv/\sqrt{2}$ everywhere,
which has $\xi=0$ for $r<r_1$ and has $\xi=2\pi$ for $r>r_2$
and in which $\xi$ changes smoothly from $0$ to $2\pi$ for
$r_1<r<r_2$. For this configuration to unwind, either $\rho$
or $\sigma$ must vanish, but this costs only a finite energy.
Indeed from (\ref{Energy}), we note that at large radius $\rho$
can vanish at small cost in energy, while at small radius $\sigma$
can vanish at small cost in energy. Furthermore, because the
$\xi=0$ and $\xi=2\pi$ vacua are in fact the same configuration,
no fields need be changed for $r$ outside $r_1<r<r_2$ in order
to unwind $\xi$. We conclude that the theory does not have
topological solitons of this kind.

\vskip .3in

\section{The Amplitude Expansion}
\label{sec4}

In this section, we discuss ``typical'' solutions
to the equations of motion in which the amplitudes
of the gauge invariant fields (and consequently
the energy density) are arbitrarily small at
arbitrarily early and late times.
Not all solutions exhibit this behavior.
For example, the sphaleron is a static solution
and therefore the magnitudes of the fields are
constant in time.  One can also imagine solutions
which are asymptotically equal to the sphaleron for
early (late) times
but which dissipate into small amplitude configurations
at late (early) times.
Thus, by restricting ourselves to solutions which
dissipate both in the future and the past, we are
excluding some solutions from our treatment.
For the solutions we wish to treat, at both
early and late times it is appropriate to
expand the equations of motion as power
series in the amplitudes of the fields.
At sufficiently early and late times, we
need only consider the lowest order (linear)
equations whose solutions in fact do dissipate
both in the past and the future.

The amplitude expansion is equivalent to
a coupling constant expansion, as we now show.
First, it is necessary to restore the factors
of $g$ which were scaled out in
(\ref{SphAn}). This is done by replacing
$\psi$, $\rho$, $\sigma$, and $\xi$ by
$g \psi$, $g \rho$, $g\sigma$ and
$g \xi$. The vacuum is now given by $\rho=1/g$, $\sigma=v/\sqrt{2}$,
$\psi=0$ and $\xi=0$, and it is convenient to define the
new parameters
\begin{equation}
m=\frac{1}{2}gv ~~~~~~ m_H=\sqrt{2\lambda}v ~~~~~~
\bar\lambda=\frac{\lambda}{g^2} \ ,
\label{newparams}
\end{equation}
and to define the
shifted fields  $y$
and $h$ by
  \begin{mathletters}
  \begin{equation}
  g\,\rho(r,t) = 1 + g\,y(r,t)
  \label{rhoShift}
  \end{equation}
  \begin{equation}
  g\,\sigma(r,t) =  \sqrt{2} m + \frac{g\,h(r,t)}{r} \ .
  \label{sigmaShift}
  \end{equation}
  \label{varShift}
  \end{mathletters}
We wish to to expand the equations of motion order by
order in the fields $y$, $h$, $\psi$ and $\xi$, all of which
vanish in vacuum.  This is equivalent to performing a power
series expansion in $g$ with $\bar\lambda$, $m$, and $m_H$
held fixed, because upon making the substitutions (\ref{varShift})
and (\ref{newparams}) in the equations of motion, every $y$,
$h$, $\psi$ or $\xi$ is multiplied by a single $g$, and no other
$g$'s occur. In doing an amplitude expansion, it is very helpful
to have the equations of motion written in terms of gauge invariant
variables.
If we were using gauge variant variables,
a large amplitude field could carry zero energy.
In the formulation we are using, the vacuum is a single
point in the configuration space  of the gauge invariant
variables, and perturbations described by nonzero $y$,
$h$, $\psi$, or $\xi$ must carry energy.

We now expand each of the fields $y$, $h$, $\psi$ and
$\xi$ in powers of $g$, expand the equations of motion
in $g$, and keep only the terms linear in $g$. The linearized
equations are
\smallskip
  \begin{mathletters}

  \begin{equation}
  \Biggl(\partial^\mu\partial_\mu   - m^2   -\frac{2}{r^2}\Biggr)\,
  y =0
  \label{giRhoLin}
  \end{equation}

  \begin{equation}
  \Biggl(\partial^\mu\partial_\mu  - m_H^2 \Biggr)\,h   =0 \
  \label{giSigmaLin}
  \end{equation}
\label{linkappataueqs}
\end{mathletters}
\begin{mathletters}

  \begin{equation}
  \partial^\mu\left\{ \frac{\partial_\mu  \psi +  \epsilon_{\mu\nu}
  \partial^\nu  \xi}{1 + \frac{1}{2}r^2m^2}\right\} - \frac{2}{r^2} \,  \psi =
0
 \
  \label{giPsiLin}
  \end{equation}
  \begin{equation}
  \partial^\mu\left\{ \frac{  \frac{1}{2}  r^2m^2 \partial_\mu  \xi -
  \epsilon_{\mu\nu}\partial^\nu \psi  }{1 + \frac{1}{2}r^2m^2 }
  \right\} - m^2 \xi= 0  \ .
  \label{giXiLin}
  \end{equation}
  \label{linpsixieqs}
  \end{mathletters}
We have uncoupled equations for $y$ and $h$
while the $\psi$ and $\xi$ equations remain coupled.
The field $y$ satisfies the equation of motion
for the angular momentum $l=1$
partial wave of a
field with the $W$ boson mass $m$, and the field
$h$ satisfies the equation of motion for
the $l=0$
partial wave of a
field with the Higgs mass $m_H$.

Now, let us turn to decoupling equations
(\ref{linpsixieqs}).
Define the new variables $x(r,t)$ and $z(r,t)$ through
\begin{mathletters}
\begin{equation}
x= -\frac{2}{r}\psi + \frac{2}{1+\frac{1}{2}r^2m^2}
\Biggl(\frac{1}{2}r^2m^2\dot\xi
- \psi' \Biggr)
\label{xdef}
\end{equation}
\begin{equation}
z= \frac{2}{r}\psi + \frac{1}{1+\frac{1}{2}r^2m^2}
\Biggl(\frac{1}{2}r^2m^2 \dot\xi - \psi' \Biggr)  \ ,
\label{zdef}
\end{equation}
\label{xzdefs}
\end{mathletters}
where $\dot{} =\partial/\partial t$ and $' = \partial/\partial r$.
The linear equations (\ref{linpsixieqs})
for $\xi$ and $\psi$ are equivalent to the decoupled
equations
\begin{mathletters}
\begin{equation}
\Biggl( -\partial^\mu\partial_\mu + m^2 \Biggr)\,x = 0
\label{yeqn}
\end{equation}
\begin{equation}
\Biggl(-\partial^\mu\partial_\mu + m^2 + \frac{6}{r^2} \Biggr)\,z = 0 ~.
\label{zeqn}
\end{equation}
\label{yzeqns}
\end{mathletters}
We see that $x$ and $z$ satisfy
the equations of motion for
$l=0$ and $l=2$ partial waves of a field with mass $m$.

To better understand the angular momentum decomposition
it is convenient to return to (\ref{SphAn}),
set $g=1$,
write $A_i^b$
in terms of gauge variant variables as
\begin{equation}
A_i^b = \Bigl( 2\alpha + r a_1 \Bigr)
\frac{\delta_{ib}}{3r}\, -\, \gamma\,\frac{\epsilon_{ibk}\hat x^k}{r}
\,-\, \Bigl(\alpha- r a_1 \Bigr)\Biggl(
\frac{3\hat x_i \hat x_b - \delta_{ib}}{3r}\Biggr) \ ,
\label{pwaves}
\end{equation}
and work in $A_0=0$ gauge.
Note that $A_i^b$ is the sum of terms with $j=0,1,$ and $2$ where
$j$ is the sum of isospin, orbital angular momentum and spin.
Working to linear order in $\alpha$, $\gamma$, and $a_1$, we
now relate the terms appearing in (\ref{pwaves}) to
the variables $x$, $y$, and $z$.  First, from the definitions (\ref{defchi}),
(\ref{rhoDef}), and (\ref{rhoShift})
we see that to linear order,
$\gamma=-y$.
Next, from (\ref{psiDef}), and linearizing (\ref{AAA}),
we obtain
\begin{equation}
\dot a_1 = -\,\frac{2}{r^2}\,\psi
\label{dota1}
\end{equation}
\begin{equation}
\dot\alpha = \frac{ \frac{1}{2}r^2m^2\dot\xi - \psi' }
{1+\frac{1}{2}r^2m^2} \ .
\label{dotalpha}
\end{equation}
Comparing with the definitions (\ref{xzdefs}),
we find
\begin{mathletters}
\begin{eqnarray}
\label{xrel}
 2\dot \alpha + r \dot a_1 &=& x \\
\label{yrel}
 -\gamma &=& y \\
\label{zrel}
 \dot \alpha - r \dot a_1 &=& z\ .
\end{eqnarray}
\end{mathletters}
Recall that the variables $x$, $y$
and $z$ satisfy the equations of motion for the partial
waves of a field with mass $m$ and angular momentum
$l=0,1,$ and $2$ respectively. We now see that they are also associated
with $j=0,1,$ and $2$ respectively.  Thus, these modes behave as if their
angular momentum $l$ is determined by the sum of their
orbital angular momentum, spin and isospin\cite{jacreb}.

Before solving the linearized equations
(\ref{linkappataueqs}) and (\ref{yzeqns}), we
must specify the $r=0$ boundary conditions.
Using the boundary conditions on $\psi$,
$\rho$, $\sigma$, and $\xi$ presented after
(\ref{giEqs}), we see that $x=y=z=h = 0$
at $r=0$. The initial value data given as
$\rho\, , \, \sigma\, , \, \psi$, $\xi$ and their
time derivatives
at some time are equivalent to initial value
data for $x$, $y$, $z$, $h$, and their time
derivatives at that time.  (From (\ref{xzdefs}),
we note that determining the  initial values
of $\dot x$ and $\dot z$ requires knowledge
of the initial value of $\ddot \xi$.  This can
be obtained from the other initial data using
(\ref{giXiLin}).) After solving (\ref{linkappataueqs})
and (\ref{yzeqns})
for $x$, $y$, $z$, and $h$
as we describe below, we can use
(\ref{varShift}) and
\begin{mathletters}
\begin{equation}
\psi=\frac{r}{6}( 2z-x )
\label{psifromxz}
\end{equation}
\begin{equation}
\dot \xi = \frac{\psi' + \frac{1}{3} \left( x+z \right)
\left(1+\frac{1}{2}r^2 m^2 \right) }{\frac{1}{2} r^2 m^2}
\label{xifromxz}
\end{equation}
\label{psixifromxz}
\end{mathletters}
obtained from (\ref{xzdefs})
to determine the solutions (to lowest order in $g$) for
$\rho$, $\sigma$, $\psi$, and $\dot\xi$.
Finally, the solution for $\xi$
can be obtained from that for $\dot\xi$ using the
initial data for $\xi$.

We must solve the equation
\begin{equation}
\Biggl(\partial_t^2 - \partial_r^2 + m^2 +
\frac{l(l+1)}{r^2}\Biggr)F_l(r,t)=0
\label{Feq}
\end{equation}
subject to the boundary condition $F_l(0,t)=0$.
We will then use
$l=0,1,$ and $2$ solutions for $x,y,$ and $z$,
and an $l=0$ solution with $m$ replaced by $m_H$ for $h$.
The general solution to (\ref{Feq}) is
\begin{equation}
F_l(r,t)= \int_0^{\infty} \frac{ {\rm d}k}{2\pi} \Biggl\{
f_l(k)\,kr\,j_l(kr) \exp (-i\omega_k t  ) ~+~{\rm c.c.} \Biggr\}
\label{feq}
\end{equation}
where $\omega_k=\sqrt{k^2+m^2}$ and $j_l(s)$ satisfies
\begin{equation}
\Biggr(\frac{{\rm d}^2}{{\rm d}s^2} + 1 - \frac{l(l+1)}{s^2}
\Biggl) \,s\,j_l(s) = 0
\label{geq}
\end{equation}
with $\lim_{s\rightarrow 0}s\,j_l(s)=0$.
The solutions we require are:
\begin{mathletters}
\begin{eqnarray}
s\,j_0(s) &=& \sin s \\
s\,j_1(s) &=& -\cos s + \frac{ \sin s }{s} \\
s\,j_2(s) &=& -\sin s - \frac{ 3 \cos s}{s} + \frac{3\sin s}
{s^2} \ .
\label{gsolns}
\end{eqnarray}
\end{mathletters}
Finally, the complex function $f_l(k)$ can be determined
from the initial conditions on $F_l(r,t)$ and $\dot F_l(r,t)$.
So, given initial conditions on $x$, $y$, $z$,
and $h$ and their time derivatives,
we can now obtain explicit solutions to
the linearized equations of motion.

The solutions to the linearized equations look like shells of
energy which come in from large $r$, bounce, and then recede
to large $r$ again.  Because of the mass term in (\ref{Feq}) the
shells do not maintain their shape as they propagate at large
$r$. As $t\rightarrow \pm \infty$ the solutions disperse and the
amplitudes of the fields decrease.  This means that for the solutions
we are discussing, using the linearized equations is a better and
better approximation at larger and larger $|t|$. In the far past, solutions
to the full nonlinear equations of motion are well approximated by
solutions to the linear equations specified by the functions
$f_l^p(k)$, and in the far future they are once again well
approximated by solutions to the linear equations specified by
different functions $f_l^f(k)$.   Working order by order in the
amplitude expansion, one could obtain $f_l^f$ perturbatively
given $f_l^p$.

\vskip .3in

\section{${\pmb Q}$, ${\pmb\Delta}{\pmb N}_{\pmb H}$, and the Sphaleron}
\label{sphalsection}

In the spherical ansatz, the topological charge
  \begin{equation}
   Q = \frac{g^2}{32\pi^2} \int_{-\infty}^\infty dt  \,
  \int d^3 x \,
  \epsilon^{\mu\nu\alpha\beta}
  {\rm Tr}(F_{\mu\nu} F_{\alpha\beta}) \ .
  \label{QDef}
  \end{equation}
can be written \cite{W}
  \begin{equation}
   Q =   -\frac{1}{2\pi}\int_{-\infty}^\infty dt  \,
  \int_0 ^{\infty}dr \,  \epsilon^{\mu\nu}  \partial_\mu
  \left[ a_\nu +\frac{i}{2}\,  \left(\chi D_\nu\chi^*  - \chi^*
  D_\nu\chi \right)  \right]    \ .
  \label{Qspherical}
  \end{equation}
The integrand is the divergence of a gauge variant
current, and so cannot be written
in terms of gauge invariant variables.
For solutions to the equations of motion,
however, we can use equations (\ref{rhoDef}),
(\ref{psiDef}),
(\ref{AAA}), and (\ref{giPsi}) to write a gauge invariant current
  \begin{equation}
   k^\mu(r,t) = -\frac{1}{2\pi} \,  \left[  \frac{
   (1-\rho^2)( \partial^\mu\psi - \frac{1}{4}r^2\sigma^2
   \epsilon^{\mu\nu}\partial_\nu\xi )}
  {(\rho^2 + \frac{1}{4}r^2\sigma^2)}
     \right]
     -\frac{1}{2\pi} \epsilon^{\mu\nu} \partial_\nu \xi\ ,
  \label{kDef}
  \end{equation}
which is zero in vacuum regions of space-time and
which satisfies
  \begin{equation}
  \frac{g^2}{8\pi}\, r^2\,
  \epsilon^{\mu\nu\alpha\beta}
  {\rm Tr}(F_{\mu\nu} F_{\alpha\beta} ) = \partial_\mu k^\mu \ .
  \label{Qgi}
  \end{equation}
It is shown in Ref. \cite{fiveofus}, without reference to the spherical
ansatz, that the topological charge of a solution to the equations of
motion is not well-defined as an integral over all space-time.
Attempting to evaluate $Q$ as the limit of a sequence
of integrals taken over larger and larger regions
of space-time such that in the limit
all of space-time is included, one can obtain different results
for different sequences of regions.  This observation applies
to all solutions which linearize in the far future or in the far past.
Thus although one could obtain a finite result upon
evaluating (\ref{Qspherical}) for a solution, we do not
believe that the result would have any significance.

{}From Ref. \cite{fiveofus}, we learn that it is
profitable to characterize
solutions
by their change in
Higgs winding number.
The matrix $\Phi$ of (\ref{higgsmatrix}) can be
written in terms of a unitary matrix $U$ according to
\begin{equation}
\Phi({\bf x},t)
= (\varphi_1^*\varphi_1 +
\varphi_2^*\varphi_2 )^{1/2} U({\bf x},t)
\label{Udef}
\end{equation}
at all space-time points where $|\varphi |\equiv (\varphi_1^*\varphi_1 +
\varphi_2^*\varphi_2)^{1/2}$ is non-vanishing.
We will consider only those solutions for which the
fields approach their vacuum values in the $|{\bf x}|\rightarrow\infty$
limit for all $t$.
Without loss of generality, we can work in a gauge in which the
boundary condition
\begin{equation}
\lim_{|{\bf x}|\rightarrow\infty}U({\bf x},t) = 1
\label{Ubc}
\end{equation}
is satisfied.
At any time $t$ when $|\varphi|\neq 0$ throughout
space, the configuration can be characterized
by the integer valued winding number
\begin{equation}
N_H(t)=\frac{1}{24\pi^2}\int {\rm d}^3x \epsilon^{ijk}
{\rm Tr}\biggl( U^\dagger \partial_i U \,
U^\dagger \partial_j U \,U^\dagger \partial_k U \biggr) \ .
\label{NHdef}
\end{equation}
The Higgs winding number $N_H$ is gauge invariant
under small gauge transformations, but large
gauge transformations can change it by an integer.
For the solutions of
interest to us, at early and late times the
fields are close to their vacuum values.
Therefore, $N_H(t)$ becomes constant in time in the
far past and in the far future and we can define the gauge
invariant quantity
\begin{equation}
\Delta N_H = \lim_{t\rightarrow\infty}N_H(t)
- \lim_{t\rightarrow -\infty}N_H(t) \ .
\label{DeltaNHsphandef}
\end{equation}

We now state the results concerning
$\Delta N_H$ proved in Ref. \cite{fiveofus}.
These results were established
without reference to the spherical ansatz.
First,
solutions with $\Delta N_H \neq 0$ which dissipate
at early and late times
must have at least
the sphaleron energy.
Second,
suppose we couple a quantized chiral fermion to
the classical gauge and Higgs backgrounds
considered in this paper, giving the fermion
a gauge invariant mass via a Yukawa coupling.
The number of fermions produced in a background
given by a solution which dissipates at early and
late times is equal to $\Delta N_H$\cite{ghk}.

In the spherical ansatz, $\Phi$ is given by
\begin{equation}
\label{Phisph}
\Phi({\bf x},t) = \frac{1}{g}\, \sigma (r,t)
\exp \left[ i \eta(r,t)
{\pmb\sigma}\cdot{\bf\hat x} \right]
\end{equation}
where the boundary
condition (\ref{Ubc}) on $U$ can be implemented by
working in a gauge in which
\begin{equation}
\lim_{r\rightarrow \infty}\eta(r,t) = 0 \ .
\label{etabc}
\end{equation}
The boundary condition (\ref{firstetabc}), namely that
$\eta(0,t)$ is an integer multiple of $\pi$, ensures that
(\ref{Phisph}) makes sense at the origin.
At times when $\sigma \neq 0$ for all $r$,
the Higgs winding number (\ref{NHdef}) is given by
\begin{equation}
N_H(t)
= \frac{1}{\pi} \int_0^\infty {\rm d}r
\frac{\partial\eta}{\partial r}\left(1- \cos 2 \eta \right)
= -\frac{\eta(0,t)}{\pi}  \ ,
\label{NUsphandef}
\end{equation}
where we have used both (\ref{etabc}) and (\ref{firstetabc}).
At times when $\sigma$ is nonzero at the origin, $\eta(0,t)$
is constant and therefore so is $N_H(t)$.  The Higgs
winding number can change only at times when $\sigma$
vanishes at $r=0$.

$\Delta N_H$, unlike $N_H(t)$, is gauge invariant and
we now show how to write it in terms of gauge invariant
variables in the spherical ansatz.  We assume that zeros
of $\sigma$ occur only at isolated points  in the \hbox{$(r,t)$}
plane. For $\sigma=(\mu^2+\nu^2)^{1/2}$ to vanish,  both
$\mu$ and $\nu$ must vanish at the same \hbox{$(r,t)$}  point.
Typically, the zeroes of $\mu$ and $\nu$ each  define one
dimensional curves that  intersect at isolated points.  So we
believe that the case we are treating, for which the zeros of
$\sigma$ are isolated, is generic. {}From (\ref{DeltaNHsphandef}),
we can express the change in Higgs winding as
\begin{equation}
\label{contour}
\Delta N_H = -\,\frac{1}{\pi} \int_C {\rm dx^\mu}
\partial_\mu\eta(r,t)\ ,
\end{equation}
where the contour $C$ is oriented from the infinite past to
the infinite future, following $r=0$ except for infinitesimal
excursions to non-zero $r$ to  avoid any zeroes of  $\sigma$
which lie along $r=0$.  In this way we ensure that in the generic
case all other zeros of $\sigma$ lie to the right of $C$,  since
in this case  the zeros of $\sigma$ are isolated.  Furthermore,
all zeros of $\rho$ lie to the right of $C$,  since
$\rho$ is  continuous and equal to unity along  $r=0$.  Since
neither $\sigma$ nor $\rho$ vanish on $C$, both $\eta$
and $\theta$ are defined on $C$, and therefore $\xi=
\theta-2\eta$ is also defined on $C$. From  the boundary
condition  (\ref{firstthetabc}) on $\theta$, we see that equation
(\ref{contour}) implies
\begin{equation}
\label{gideltanh}
\Delta N_H = \,\frac{1}{2\pi} \int_C {\rm dx^\mu}
\partial_\mu\xi(r,t)\ .
\end{equation}
Note that (\ref{gideltanh}) is manifestly gauge invariant because
$\xi$ is gauge invariant. Furthermore, expression  (\ref{gideltanh})
is valid for any finite energy sequence of configurations parameterized
by $t$ whose zeros of $\sigma$ are isolated and whose zeros of
$\sigma$ at $r=0$ are restricted to a finite range of $t$, and not just
for sequences which are solutions to the equations of motion.

Solutions with $\Delta N_H \neq 0$ must have at least
one time when $\sigma(0,t)=0$. (Even outside the spherical
ansatz, the Higgs field must vanish at some point in space-time
in order for the Higgs winding to change.) We now show
that solutions in the spherical ansatz with $\Delta N_H \neq 0$
must have $\rho = 0$ at some point in the $(r,t)$ plane.  Because
we are discussing solutions with $\xi\rightarrow 0$ for $r\rightarrow
\infty$, in a gauge in which (\ref{etabc}) is satisfied we also have
\begin{equation}
\lim_{r\rightarrow \infty}\theta(r,t) = 0\ .
\label{thetabc}
\end{equation}
For the solutions we are considering, the amplitudes
of all the fields dissipate at early and late times, and in
particular $\xi=\theta-2\eta$ dissipates. This means that
\begin{equation}
\lim_{t\rightarrow\pm\infty} \int_0^\infty {\rm dr}
\frac{\partial\xi}{\partial r} = 0 \ ,
\label{xiwinding}
\end{equation}
which in turn implies that
\begin{equation}
\lim_{t\rightarrow\infty}\frac{1}{2\pi}\int_0^\infty {\rm dr}
\frac{\partial\theta}{\partial r}
- \lim_{t\rightarrow -\infty}\frac{1}{2\pi}\int_0^\infty {\rm dr}
\frac{\partial\theta}{\partial r} = \Delta N_H \ .
\label{thetawinding}
\end{equation}
Together with the conditions (\ref{firstthetabc}) and  (\ref{thetabc}),
equation (\ref{thetawinding}) implies that $\theta$ changes by
$2\pi \Delta N_H$ as one traverses a large rectangular path in the
\hbox{$(r,t)$} plane with sides at $r=0$ and at large $r$, and
constant $t$ sides in the far future and far past.  Hence,  for a solution
to have nonzero $\Delta N_H$, there must be at least one point  in
the $(r,t)$ plane where $\rho$ vanishes.

The sphaleron configuration of Manton and Klinkhamer
\cite{manton} is given in terms of our gauge invariant
variables by
  \begin{mathletters}
  \begin{eqnarray}
  \rho^{\rm sph}  &=&  |2 f(r) -1|  \\
  \sigma^{\rm sph}  &=& \frac{gv}{\sqrt{2}}\, h(r)    \\
  \psi^{\rm sph} &=& 0 \\
  \xi^{\rm sph} &=& \cases{\pi~; & $0<r<r_0$ \cr 0~; & $r>r_0$ \cr } \,
  \end{eqnarray}
  \label{gisphal}
  \end{mathletters}
where $f(r)$ and $h(r)$ are continuous functions that vary from
zero to one as $r$ increases from zero to infinity. As $f(r)$ varies
between zero and one, there is a nonzero radius $r_0$ for which
$f(r_0)=1/2$.  On this shell, $\rho$ vanishes and $\theta$ is not
defined; whereas $\eta$ is not defined at $r=0$, where $\sigma$
vanishes. The variable $\xi$ is not defined at both $r=0$ and $r=r_0$.
In general, a solution with $\Delta N_H \neq 0$ can have
its zeros of $\rho$ and $\sigma$ at different times.
In the special case where such a solution goes exactly
through the sphaleron configuration at some time $t_0$, then
it has both a zero of $\rho$ and of $\sigma$ at that time.
In such a case, in the $(r,t)$ plane $\xi$ winds by $2\pi$
in one direction about $(0,t_0)$ and in the opposite
direction about $(r_0,t_0)$.

\section{Conclusions}
\label{sec6}

We have cast the full nonlinear equations of motion
in the spherical ansatz as four equations for four
gauge invariant degrees of freedom in such a way that
the Gauss's law constraint can easily be implemented.
We hope these equations are of use for numerical
study or for those seeking exact solutions. We have
also studied the linear form of these equations, which
are the leading term in an amplitude expansion.
The linear equations we obtain describe an angular
momentum $l=0$ partial wave of a field with the Higgs
mass, and angular momentum $l=0,1,$ and $2$ partial
waves of a field with the $W$ boson mass.  These modes
behave as if  their angular momentum $l$ is determined by
the sum of their orbital angular momentum, spin and isospin.
Typical solutions to the full equations of motion are well
approximated by solutions to the linearized equations at early
and late times. Furthermore, the fields approach their vacuum
values and the Higgs winding number is constant in the
far past and in the far future. The change in Higgs winding,
which is related to anomalous fermion production
\cite{fiveofus,ghk},  can be expressed as an integral involving
the gauge invariant variables.

\acknowledgments

We would like to thank J. Goldstone and C. Rebbi for very helpful
discussions. We particularly want to thank C. Rebbi for help in
decoupling equations (4.4), and for helping us to understand
the boundary conditions at the origin.

\end{document}